\documentclass[pre,preprint,aps,eqsecnum]{revtex4}
\usepackage{graphicx}

\begin{document}

\title{Dynamics and Thermodynamics of the Glass Transition}

\author{J. S. Langer}

\affiliation{Department of Physics,
University of California,
Santa Barbara, CA  93106-9530  USA}

\date{January, 2006}

\begin{abstract}
The principal theme of this paper is that anomalously slow, super-Arrhenius relaxations in glassy materials may be activated processes involving chains of molecular displacements.  As pointed out in a preceding paper with A. Lemaitre, the entropy of critically long excitation chains can enable them to grow without bound, thus activating stable thermal fluctuations in the local density or molecular coordination of the material. I argue here that the intrinsic molecular-scale disorder in a glass plays an essential role in determining the activation rate for such chains, and show that a simple disorder-related correction to the earlier theory recovers the Vogel-Fulcher law in three dimensions. A key feature of this theory is that the spatial extent of critically long excitation chains diverges at the Vogel-Fulcher temperature. I speculate that this diverging length scale implies that, as the temperature decreases, increasingly large regions of the system become frozen and do not contribute to the configurational entropy, and thus ergodicity is partially broken in the super-Arrhenius region above the Kauzmann temperature $T_K$. This partially broken ergodicity seems to explain the vanishing entropy at $T_K$ and other observed relations between dynamics and thermodynamics at the glass transition.
\end{abstract}

\maketitle

\section{Introduction}

In a preceding paper \cite{LL05}, Lemaitre and I explored the hypothesis that anomalously slow,  super-Arrhenius, relaxation rates in glassy solids can be understood by assuming that transitions between the inherent states of such materials are enabled by thermally activated chains of small molecular displacements.  (See \cite{JAMMING} for a summary of research in a wide range of topics related to the dynamics of glassy materials.) More specifically, we developed a model of the spontaneous formation of shear-transformation-zones (STZ's) by thermal fluctuations in the absence of driving forces. STZ's are localized irregularities in the density of molecules, or in near-neighbor molecular correlations, that undergo irreversible rearrangements during shear deformation.\cite{FL98,FLP04,JSL04}  We visualized their formation as the glassy analog of the formation of vacancy-interstitial pairs.  At low temperatures, such pairs must become well separated in order to be thermodynamically stable against recombination; and the state that includes such a well separated pair is inherently distinct from the state in which the pair is absent.  In our model, the activation energy for forming a stable pair is the free energy -- including the entropy -- of a chain of molecular displacements that moves the ``interstial'' away from the ``vacancy'' and is just long enough to be marginally unstable against further growth. In short, we solved a nucleation problem in which the relevant reaction coordinate is the length of the chain.  The entropy associated with different chain configurations is a measure of the number of routes across the activation barrier (see \cite{JSL69}), and therefore plays a central role in determining the transition rate.  

The analysis presented in \cite{LL05} was at best only partly successful.  The excited chains seemed to have qualitatively the right properties to predict a diverging viscosity at a non-zero temperature $T_0$.  In order to produce a Vogel-Fulcher law, however, we had to assume that the chains were restricted to lie on two-dimensional surfaces, perhaps the interfaces beween the frustration-limited domains of Kivelson {\it et al} \cite{KIVELSON95}, or boundaries within the mosaic structures proposed by Kirkpatrick, Thirumalai, Wolynes and others.\cite{WOLYNES89,WOLYNES00,WOLYNES05}  This picture may indeed deserve further investigation; I shall return to it briefly in discussing the thermodynamics of these systems.  The main purpose of the present paper, however, is to argue that the missing feature of the model discussed in \cite{LL05} may be the spatially disordered environment that is intrinsic to any glassy material.  

The problem of computing the activation energy of excitation chains is related -- but not exactly equivalent -- to the problem of computing the statistics of self-avoiding random walks. That problem, in turn, is approximately equivalent to solving a diffusion equation in a self-consistent repulsive potential.  In \cite{LL05}, we approximated the solutions of those related problems by adapting Flory's method for calculating excluded-volume effects in polymers.\cite{FLORY53}  I do essentially the same thing here, but solve a (non-conserving) diffusion equation in a random potential whose spatial disorder is that of the configurational degrees of freedom of the glass.  It is well known that diffusion is constrained in disordered systems; thus the statistics of excitation chains must be determined by the interplay between two competing effects: swelling of the excitation region due to excluded volume, and contraction due to disorder.  The result of this competition is the Vogel-Fulcher law in three dimensions. 

Before entering into the details of this analysis, a few remarks about how it is related to previous work in this field are in order.  The scientific literature abounds with attempts to explain super-Arrhenius behaviors in glasses.  Some of the most important and influential of these papers, notably that of Adam and Gibbs \cite{ADAM-GIBBS65}, are almost half a century old.  More recent work along these lines has been carried out by Wolynes and his collaborators \cite{WOLYNES89,WOLYNES00,WOLYNES05} and by other investigators as reviewed in \cite{TARJUS-KIVELSON}. (See \cite{BOUCHAUD04} for a perceptive analysis of the ``Adam-Gibbs-Kirkpatrick-Thirumalai-Wolynes scenario.'')  All of these authors make one fundamental assumption with which I concur -- that the super-Arrhenius relaxation processes are intrinsically nonequilibrium phenomena. By this I mean that the super-Arrhenius formula pertains strictly to the rate at which a glassy material makes transitions between its inherent states \cite{GOLDSTEIN69,STILLINGER-WEBER82,STILLINGER88}, and not just to the statistics of those states themselves. This assumption sets us apart, for example, from Cohen and Grest who, in their classic paper \cite{COHEN-GREST}, attributed super-Arrhenius behavior to percolation of liquid-like regions in equilibrated states.  An especially relevant counterexample is the long-standing convention in engineering-oriented papers on metallic glasses and other amorphous systems, where it is assumed that the equilibrium density of flow defects is a super-Arrhenius function of temperature.  For example, see \cite{SPAEPEN77,ARGON79,DUINE,TUINSTRA,DEHEY,WLJ05}. I shall not dwell on this issue, but it needs to be taken into account when reading that literature.  

The principal way in which the super-Arrhenius processes are relevant to thermodynamic equilibrium is that they determine the validity of the ergodic hypothesis.  It is in this connection that I disagree with the approach of Wolynes {\it et al}, which is based on the idea that the activated transition states consist of droplets of an entropically favored liquid phase.  Such a phase seems to be precluded by the same, extremely slow relaxation rates that all of these authors are trying to compute; it cannot exist on  experimentally relevant time scales, especially not as an ephemeral transition state.  Rather, the activated chains discussed in \cite{LL05} and in the present paper might provide a better model of the transition mechanism because they are ordinary thermal fluctuations occurring within single inherent states.  Moreover, the excitation-chain model involves only short-ranged interactions between the constituent molecules; it draws none of its ingredients from infinite-range mean-field models or the like. Thus, at least in principle, it should be possible to use it to develop microscopic descriptions of various kinds of glassy systems. 

The scheme of this paper is as follows.  Section II contains a summary of  basic ideas and principal predictions of the excitation-chain dynamics.  Section III is devoted to the mathematics of the disorder problem in glassy systems and the derivation of a formula used in Section II. Readers who are not interested in such details may skip that part of the paper; but the analysis described there addresses some technical issues that distinguish this nonequilibrium situation from otherwise similar problems. In Section IV, I propose an interpretation of the preceding results that seems to explain the striking relationships that have been found to exist between the dynamics and thermodynamics of the glass transition -- that is, the apparent equivalence of the Vogel-Fulcher and Kauzmann temperatures, the success of the Adam-Gibbs theory, and the approximate proportionality between the dynamic fragility and the jump in the specific heat at the glass temperature. I close with a list of unanswered questions.

\section{Excitation-Chain Dynamics} 

All of the discussion that follows is based on a picture of a glass as a supercooled liquid in which configurational rearrangements have become very much slower than the thermal vibrational motions of the molecules within their local environments, {\it i.e.} within their ``cages.'' The goal of the excitation-chain theory is to compute the rates of configurational rearrangements, and to do this using a mechanism that involves just the rapid thermal fluctuations that ultimately must drive those motions.  An essential element of this picture is the notion of frustration -- that the energetically most preferred local configurations of the molecules do not fit together to fill space, and therefore that there is a geometrically necessary population of somewhat higher energy ``defects'' where the local coordinations are not the most preferred ones.  We may think of the distribution of these relatively populous, necessary defects as merging, at yet higher energies, with a much smaller population of shear-transformation zones (STZ's) and other localized irregularities in the density and/or molecular arrangements.  The anomalously loose STZ-like defects apparently govern the response of the system to external driving forces.  They are the source of the observed super-Arrhenius behaviors, and therefore are of special interest here.  

The relevance of one-dimensional excitations to the nonequilibrium behavior of glassy materials is supported by the molecular dynamics simulations of Glotzer and colleagues, \cite{GLOTZER99,GLOTZER00a,GLOTZER00b} which showed  that transitions between inherent states in glass-forming liquids take place via motions of stringlike groups of molecules.  The well documented existence of force chains in granular materials \cite{JAMMING} seems to be further evidence in favor of the idea that forces and displacements are transmitted primarily along one dimensional structures in noncrystalline systems. 

As in \cite{LL05}, consider just the spontaneous creation of an STZ, that is, its creation due to a thermal fluctuation in the absence of external driving; and assume that such an event is roughly similar to the formation of a ``vacancy-interstitial'' pair, followed by displacements of atoms along an excitation chain. Lemaitre and I visualized this chain as a thermal fluctuation in which a linear array of momentarily loosened atoms undergoes small displacements, effectively moving the ``vacancy'' and the ``interstitial'' far enough away from each other that they do not quickly recombine.  More precisely, the transition state for this activated process is a momentary thermal excitation of the system that enables a chain of molecular displacements just long enough that it is as likely to grow as to decay.  No molecule has yet moved fully out of its cage in this transition state. Excitation chains smaller than the critical size, like subcritical liquidlike clusters in a supercooled vapor,  with high probability just disappear, leaving the system unchanged.  Once the chain exceeds this critical size, however, the vacancy and the interstitial become uncorrelated with each other, and each finds its own stable position in a new inherent state.  The time taken by such a transition, once it occurs, is roughly the time during which the chain fluctuates between different near-critical lengths and configurations as it passes across the activation barrier. This time may be very long compared to an oscillation period for a molecule in its cage, but it is very short -- essentially instantaneous -- in comparison to the inverse of the super-Arrhenius rate at which these transitions occur.  In other words, at temperatures near the glass transition, excitation chains are very rare and relatively brief events.

The challenge is to compute the probability per unit time for formation of an excitation chain.  Consider a chain of length N, measured in units of a characteristic molecular spacing which, without loss of generality, can be set to unity; and suppose that the chain occupies a roughly spherical region of radius R in a three-dimensional space. Throughout the following, $R$ is the expected distance from the origin to the last, $N$'th link in the chain; but it seems reasonable to assume that the radius of the occupied region is the same as $R$ up to an unimportant geometrical constant.  

The required formation probability for the pair plus the excitation chain is the product of a Boltzmann factor containing the activation energy, multiplied by the number of configurations of the chain with length $N$ and extension $R$.  It is conventional to write the logarithm of this probability in the form $-\Delta G(N,R)/k_B\,T$, where $\Delta G(N,R)$ is called the activation {\it free}-energy because it includes something like an entropy.  It will become clear in Section III that this quantity is not exactly a free energy in the conventional sense.

$\Delta G(N,R)$ consists of several parts:
\begin{equation}
\label{DeltaG1}
\Delta G(N,R)= \Delta G_{\infty} + N\,e_0 + E_{int}(N,R) - k_B\,T\,\ln\,W(N,R) .
\end{equation}
The first term, $\Delta G_{\infty}$, is the bare activation energy, that is, the energy required to form the ``vacancy'' and the ``interstitial.'' $\Delta G_{\infty}$ becomes the ordinary Arrhenius activation energy at high enough temperatures, say $T > T_A$, where $N$ must vanish because the chain is no longer needed to stabilize the excitation. (See remarks at the end of this Section.) In equilibrium situations even below $T_A$,  $\Delta G_{\infty}$ is the activation energy that occurs in a Boltzmann factor for determining the population of these defects; and detailed balance requires that this Boltzmann factor be the ratio of their creation and annihilation rates.  Different processes, associated with different kinds of defects, will have different values of $\Delta G_{\infty}$.  On the other hand, in nonequilibrium situations below $T_A$, the chain dynamics near the glass transition may be much the same for different defect-related mechanisms because the chains involve large numbers of molecules.  Thus the super-Arrhenius behavior may be common to a variety of different relaxation phenomena in a single material. 

The remaining terms in Eq.(\ref{DeltaG1}) describe the excess free energy of the chain.  The second term on the right-hand side of Eq.(\ref{DeltaG1}), $N\,e_0$, is the bare activation energy of the $N$ links of the chain, unmodified by entropy or the self-exclusion effect. The average energy per link, $e_0$, is a measure of the elastic stiffness of the molecular environments, that is, the energy required to move two molecules far enough apart from one another to allow a third to pass between them.  

The third term, $E_{int}$, makes it energetically unfavorable for the links of the chain to lie near one another.  In principle, this exclusion effect might be included directly in a sum over self-avoiding random walks, perhaps using a nonperturbative method like that described by Edwards.\cite{EDWARDS65}  For present purposes, however, it seems better not to be so ambitious, especially since the exclusion forces in this case are likely to be long-ranged.  The thermal fluctuation that loosens the molecules along the chain must push molecules closer together at points away from the chain, thus producing an extended repulsion.  As a result, Flory's mean-field approximation may be more accurate here than it is for polymers.  As in \cite{LL05}, the Flory interaction energy is proportional to the square of the string density multiplied by the volume occupied by the string. That is,
\begin{equation}
\label{Eint}
E_{int}(N,R)= k_B\,T_{int}\,{N^2\over R^3},
\end{equation}
where dimensionless geometric factors have been absorbed into the definition of $T_{int}$.  Note that this approximation makes sense only in the limit of large $N$.  The exclusion effect must disappear for short chains -- an important complication that will be discussed later.

In the last term on the right-hand side of Eq.(\ref{DeltaG1}), $W(N,R)$ is a sum over chain configurations.  Evaluating $W(N,R)$ is the crux of the present analysis. In \cite{LL05}, we wrote 
\begin{equation}
\label{W0}
 \ln\,W(N,R) \approx  \nu\,N - {R^2\over 2\,N},
\end{equation}
where $\exp\,(\nu)$ is the number of choices that a walk can make at each step and $\exp\,(- R^2/2\,N)$, the free-diffusion factor (up to a normalization constant), is the {\it a priori} probability for a chain of length $N$ to occupy a region of radius $R$. We then minimized $\Delta G(N,R)$ with respect to $R$ and maximized it with repect to $N$.  That is, we found a saddle point of $\Delta G(N,R)$, which we identified as the activation energy.  In three dimensions, our result was
\begin{equation}
\label{LL3d}
{\Delta G^*(T)\over k_B}= {\Delta G(N^*,R^*)\over k_B}\approx {\Delta G_{\infty}\over k_B} +  {\rm constant} \times {T^{3/4}\,T_{int}^{1/2}\over (T-T_0)^{1/4}},
\end{equation}
where $T_0 = e_0/(\nu\,k_B)$, and ($R^*$, $N^*$) is the location of the saddle point:
\begin{equation}
R^* \propto\left({T_{int}\over T}\right)^{1/5}(N^*)^{3/5};~~~N^*\propto  {T^{3/4}\,T_{int}^{1/2}\over (T-T_0)^{5/4}}.
\end{equation}
The exponent $1/4$ in Eq.(\ref{LL3d}) was clearly too far from the Vogel-Fulcher formula to be consistent with experiment, and thus we looked at its analog in two dimensions where that exponent turns out to be unity.  (Dimensionality entered only via the interaction term.)  

The thesis here is that the missing ingredient of the preceding analysis was the intrinsically disordered, glassy environment in which the excitation chains occur.  This disorder appears in the activation energies for individual links of the chain, and the variation of these energies reflects the structural disorder in the underlying molecular configurations. To incorporate this disorder into the evaluation of $\Delta G(N,R)$, write the activation energy at the position of the n'th link of the chain, say ${\bf r}_n$, in the form $e_0 + k_B\,T\,\varphi({\bf r}_n)$, where $\varphi({\bf r}_n)$ is a dimensionless random variable.  If the disorder is uncorrelated from site to site, and $e_0$ has been chosen to be the average activation energy per link, then 
\begin{equation}
\label{phiaverage}
<\varphi({\bf r}_n)> = 0;~~~~<\varphi({\bf r}_n)\,\varphi({\bf r}_m)>= \gamma(T)\,\delta_{n,m},
\end{equation}
where the angular brackets denote a statistical average over realizations of the disorder. The function $\gamma(T)$ is the strength of the disorder associated with the geometrically necessary, frustration-induced, configurational defects discussed at the beginning of this Section.  These defects must pervade the system, and their density should not be a strongly varying function of temperature.  Let this defect density be $n_d$ per molecule, and further suppose that the variations in $e_0$ are of the order of $e_0 = \nu\,k_B\,T_0$.  Then,
\begin{equation}
\label{gammadef}
\gamma(T)=\gamma_0\,\left({T_0\over T}\right)^2;~~~~ \gamma_0 \cong \nu^2\,n_d.
\end{equation}

Equation(\ref{gammadef}) has several important implications.  Note that $\gamma(T)$ is a decreasing function of the bath temperature $T$.  This feature reflects the fact that, at lower temperatures, the chains are  more tightly constrained to lie in the minima of the random field $\varphi({\bf r}_n)$, and thus the effective coupling between the disorder and the chains is greater.  Note also that $\gamma$ may be of order unity, especially in the neighborhood of $T=T_0$.  The density of geometrically necessary defects is likely to be substantial and, at lower temperatures, the fact that the activation energy changes rapidly from site to site may invalidate the simple approximation made in Eq.(\ref{gammadef}).  If $\gamma$ is large, then a weak-coupling expansion of the kind described in Section III would not be quantitatively accurate.  However, since this theory is meant primarily to be a plausibility argument in favor of the disorder hypothesis, it seems reasonable to assume that it gives at least a qualitatively correct answer.

The crucial prediction of the disorder theory developed in Section III is that the free diffusion factor, $\exp\,(- R^2/2\,N)$ in $W(N,R)$, becomes negligable in comparison with  a localization factor $\exp\,(-\pi\,\gamma\,R/2)$ for large $N$ and $R$.  Eq.(\ref{W0}) becomes
\begin{equation}
\label{W1}
 \ln\,W(N,R) \approx  \nu\,N - {\pi\,\gamma(T)\over 2}\,R.  
\end{equation}
Inserting Eq.(\ref{W1}) into Eq.(\ref{DeltaG1}) and computing the saddle point, I find:
\begin{equation}
\label{DeltaGnew}
{\Delta G^*(T)\over k_B} = {\Delta G(N^*,R^*)\over k_B}\approx {\Delta G_{\infty}\over k_B} + 4\,\left({\pi\over 6}\right)^{3/2}\,{\gamma(T)^{3/2}\,T^{3/2}\,T_{int}^{1/2}\over \nu\,(T-T_0)},
\end{equation}
where
\begin{equation}
\label{N*}
N^*\approx 4\,\left({\pi\over 6}\right)^{3/2}\,{\gamma(T)^{3/2}\,T^{3/2}\,T_{int}^{1/2}\over \nu^2\,(T-T_0)^2},
\end{equation}
and
\begin{equation}
\label{R*}
R^*\approx\left({6\,T_{int}\over \pi\,\gamma(T)\,T}\right)^{1/4}(N^*)^{1/2}\approx 2\,\left({\pi\over 6}\right)^{1/2}\,{\gamma(T)^{1/2}\,(T\,T_{int})^{1/2}\over \nu\,(T-T_0)}.
\end{equation}
To justify neglecting the diffusion factor, note that the first expression for $R^*$ in Eq.(\ref{R*}) implies that ${R^*}^2/N^*$ is of order unity, while $\gamma\,R^*$ grows like ${N^*}^{1/2}$ for large $N^*$.  Thus, the disorder effect restores the Vogel-Fulcher result in three dimensions, and restores agreement with experiment near $T_0$ without invoking frustration-limited domains or mosaic structures.  

These results, as they stand, do not account for the transition between liquidlike and solidlike glassy behavior that occurs when the excitation chains disappear at $T=T_A$.  The fact that $T_A$ is a well defined temperature is supported both by experimental evidence such as that shown, for example, in Fig.1 of \cite{KTZ+96}, and by the excitation-chain idea itself.  At temperatures below $T_A$, a molecule that makes a thermally activated jump to a neighboring, energetically unfavorable ``interstitial'' position is most likely to jump back to its original position in its next thermally activated transition.  At higher temperatures, on the other hand, the Boltzmann probability that favors recombination can be compensated by an entropic factor of the form $\exp (\nu_1)$ ($\nu_1 \cong \nu$), which counts the number of allowed jumps that move the interstitial further away from the vacancy.  In other words, to evaluate $T_A$, we must compute the temperature at which the critical length of a chain is $N^*=1$.  Just as the exclusion energy and the disorder are relevant when $N^*$ is large, the local environment of the vacancy and interstitial must be relevant in computing $T_A$. An accurate evaluation of $E_{int}$ for small $N$ and $R$, or preferably a non-mean-field theory of the small-chain limit, will be needed in order to construct a quantitative theory of the transition between super-Arrhenius and Arrhenius behaviors.  That problem is beyond the scope of this paper.  

\section{Perturbation Theoretic Analysis of the Disorder Effect}

This Section contains a somewhat old-fashioned demonstration of how the exponential decay law in Eq.(\ref{W1}) emerges in a self-consistent perturbation-theoretic approximation.  This approximation is srictly valid only in the limit of small $\gamma$. A better calculation, using more advanced techniques, should be feasible. 

The function $W(N,R)$ in Eq.(\ref{DeltaG1}) is a sum over chain configurations.  In \cite{LL05}, Lemaitre and I approximated it by writing
\begin{equation}
\label{calGdef}
W(N,R) \approx W_0(N,R) = e^{N\,\nu}\,{\cal G}_0(N,R,0),
\end{equation}
where ${\cal G}_0(N,R,0)$ is the three dimensional diffusion kernel with a source at $R=0$:
\begin{equation}
{\cal G}_0(N,R,0) = {1\over (2\,\pi)^{3/2}}\,e^{-R^2/2\,N}.
\end{equation}
It is useful to think of this kernel as a Wiener integral over continuous paths ${\bf r}(n)$, where $n$ is a continuous variable running from $0$ to $N$, and ${\bf r}(0)=0$, ${\bf r}(N)=R$.
That is,
\begin{equation}
{\cal G}_0(N,R,0)= \int_{{\bf r}(0)=0}^{{\bf r}(N)=R}\,\delta {\bf r}(n)\, \exp\,\left[- \int_0^N  {1\over 2}\left({d{\bf r}\over dn}\right)^2\,dn\right].
\end{equation}
(See, for example, the classic review article by Gel'fand and Yaglom \cite{GELFAND-YAGLOM60}.) In this functional form, it is clear that the sum over paths is a version of the desired  sum over chain configurations.  The Gaussian exponential factor, {\it i.e.} the Wiener measure, constrains the paths to be connected, one-dimensional objects embedded in a three-dimensional space.  This continuum approximation is convenient analytically and is perfectly accurate so long as it does not make much difference that the steps along a chain have length unity (the molecular spacing) and that $dn=1$.  

One problem here is that, strictly speaking, the glassy disorder is not consistent with the continuum limit.  It requires that each unit step along a chain, say the $j$'th, have its own extra weight factor $\exp\,(-\varphi_j)$ in the sum over configurations.  In the face of this difficulty, it is easiest to solve a slightly different problem.  Let the chains remain as continuous one-dimensional objects, with no explicit links of finite length; but suppose that they exist in the presence of a random potential $\varphi({\bf r})$, defined over the whole volume of the system spanned by the continuous variable ${\bf r}$; and suppose that, in analogy to Eq.(\ref{phiaverage}),
\begin{equation}
\label{phiaverage2}
<\varphi({\bf r})> =0;~~~~<\varphi({\bf r})\,\varphi({\bf r}')>= \gamma\,\delta\,({\bf r}-{\bf r}').
\end{equation}
The assumption of delta-function (white noise) correlations for the random field $\varphi({\bf r})$, implies that the molecular length scale (unity) on which this field varies is very much smaller than any other relevant length scale in the problem, especially $R$.  Also, assume that the delta-correlations arise from a Gaussian distribution over the $\varphi({\bf r})$.  That is, the average of any function $A(\{\varphi\})$ is 
\begin{equation}
\label{functionaverage}
\langle A \rangle = \int\,\delta\varphi({\bf r})\,\exp\,\left[-{1\over 2\,\gamma}\,\int\,\varphi^2({\bf r})\,d{\bf r}\,\right]\,A(\{\varphi\}),
\end{equation}
where the functional differential $\delta\varphi({\bf r})$ is defined to include a normalization factor so that $<1>=1$. With these assumptions, replace ${\cal G}_0$ in Eq.(\ref{calGdef}) by
\begin{equation}
{\cal G}(N,R,0,\{\varphi\})= \int_{{\bf r}(0)=0}^{{\bf r}(N)=R}\,\delta {\bf r}(n)\, \exp\,\left\{- \int_0^N  \left[{1\over 2}\left({d{\bf r}\over dn}\right)^2+\varphi({\bf r})\right]\,dn\right\}.
\end{equation}
The weight function $W(N,R)$ must then be obtained by computing $\langle{\cal G}(N,R,0,\{\varphi\})\rangle$ according to Eq.(\ref{functionaverage}). 

This model is now in familiar territory. It is closely related to the quantum model for a single particle moving in a random potential.  ${\cal G}(N,R,0,\{\varphi\})$ satisfies the differential equation
\begin{equation}
\label{Gdiffeq}
\left[{\partial \over \partial N}- {1\over 2}\nabla_{\bf R}^2 +\varphi({\bf R})\right]\,{\cal G}(N,{\bf R},0,\{\varphi\})= \delta({\bf R})\,\delta(N),
\end{equation} 
Thus  ${\cal G}$ is the Green's function for the imaginary-time ($N$) Schroedinger equation for a particle moving in in a delta-correlated random potential $\varphi({\bf R})$ of strength $\gamma$.  

An apparently more closely related problem is that of polymer chains in disordered media.  In this case, the competition between strong disorder and strong self-exclusion has been explored, for example, in \cite{CATES-BALL88,GOLDSCHMIDT04}.  Lemaitre and I used the polymer analogy in our earlier paper \cite{LL05}.  This analogy does deserve to be explored further; but it may not be quite so close as it appears.  In the polymer problem, one must compute the disorder average of the thermodynamic free energy in a quenched system.  In the present case, the disorder is also quenched in the sense that it is fixed and independent of the chain configuration; but the chain is not a pre-existing entity.  The goal is to compute the disorder average of the probability that the chain appears during thermal fluctuations in the system; and thus the required quantity, $\langle{\cal G}(N,R,0,\{\varphi\})\rangle$, more nearly corresponds to an annealed average.  

The next step in this analysis is to construct a perturbation expansion for ${\cal G}$. To do this, it is easiest to work with the Laplace transform of ${\cal G}$:
\begin{equation}
\tilde {\cal G}(w,{\bf R},0,\{\varphi\})=\int_0^{\infty} dN\,e^{-w\,N} {\cal G}(N,{\bf R},0,\{\varphi\}),
\end{equation}
which satisfies
\begin{equation}
\left[w- {1\over 2}\nabla_{\bf R}^2 +\varphi({\bf R})\right]\,\tilde{\cal G}(w,{\bf R},0,\{\varphi\})= \delta(R).
\end{equation}
It also is necessary to renormalize $w$ by writing $w=w_0+w'$ and using $w_0 + \varphi$ as the perturbation.  The renormalization constant $w_0$ will be chosen as in old-fashioned particle field theory to cancel a formally (but in this case not really) divergent integral.  

Now expand $\tilde {\cal G}$ in powers of this perturbation and average the expansion term by term over $\varphi$. (The basic idea for averaging a perturbation expansion over a random potential goes back to Kohn and Luttinger \cite{KOHN-LUTTINGER57}.  That procedure, and most of the other techniques used in the next paragraphs, are described in Chapter 4 of Mahan's book on many-particle physics.\cite{MAHAN})   Averaging over the disorder restores translational symmetry, which makes it natural to work with Fourier transformed functions. Let $\hat{\cal G}(w,k)$ denote the Fourier transform of the average of $\tilde {\cal G}(w,{\bf R},\{\varphi\})$.  The calculation of $\hat{\cal G}(w,k)$ then follows completely conventional lines.  The re-summed perturbation expansion for $\hat{\cal G}(w,k)$ has the form 
\begin{equation}
\label{hatG}
\hat{\cal G}(w',k)={1\over {k^2\over 2}+w'+\Sigma(w')},
\end{equation}
where here the self-energy $\Sigma$ is a function only of $w'$.  In general, $\Sigma$ would be $k$-dependent; but that dependence would arise from short-range spatial correlations in $\varphi({\bf r})$, which are assumed to be absent. In fact, the molecular length scale, taken here to be unity, provides a necessary short wavelength cutoff.  In Eq.(\ref{hatG}), the relevant values of $k$ are of order $1/R \ll 1$, so it is appropriate (again in the limit of large $R$ and $N$) to make a small-$k$ approximation -- except, of course, when one encounters short-wavelength divergences. 

The lowest order, self-consistent approximation for $\Sigma$ is: 
\begin{equation}
\label{sigma}
\Sigma(w') \cong w_0 - \gamma\int {d^3k'\over (2\,\pi)^3}\, \hat{\cal G}(w',k'),
\end{equation}
Note that it is $\hat{\cal G}$ and not the unperturbed propagator that appears on the right-hand side of Eq.(\ref{sigma}).  This self-consistent approximation ensures that $\Sigma(w')$ and $\hat{\cal G}(w',k)$ have the same analytic structure in the complex $w'$ plane, a condition that is known to be true on very general grounds and which turns out to be essential for present purposes.   

Inserting Eq.(\ref{hatG}) into Eq.(\ref{sigma}) yields
\begin{equation}
\int {d^3k'\over (2\,\pi)^3}\, \hat{\cal G}(w',k')={k_{max}\over \pi^2} -{1\over \pi\,\sqrt{2}}\,\sqrt{w'+\Sigma(w')}.
\end{equation}
Here, $k_{max}$ is the anticipated short-wavelength cutoff that is needed in order to evaluate an $w'$-independent divergent integral over $k'$. The renormalization constant conveniently can be chosen to cancel this nominal divergence:
\begin{equation}
w_0= \gamma\, {k_{max}\over \pi^2},
\end{equation}
so that Eq.(\ref{sigma}) becomes
\begin{equation}
\Sigma(w')= {\gamma\over \pi\,\sqrt{2}}\,\sqrt{w'+\Sigma(w')},
\end{equation}
and therefore
\begin{equation}
\Sigma(w')={\pi^2\,\gamma^2\over 4}\,\left[\sqrt{1+{8\,w'\over \pi^2\,\gamma^2}}+1\right].
\end{equation}
A second shift in the Laplace variable, $w'=-\pi^2\,\gamma^2/8 +w''$ leads to
\begin{equation}
\label{hatG1}
\hat{\cal G}(w'',k)={1\over {k^2\over 2}+\left(\sqrt{w''}+{\pi\,\gamma\over 2\,\sqrt{2}}\right)^2},
\end{equation}

The last step in this part of the mathematical development is to invert the Fourier and Laplace transforms.  
\begin{eqnarray}
\label{Gfinal}
\nonumber
\langle {\cal G}(N,R)\rangle &=& e^{-N\,\Delta e_0(T)/k_B\,T}\,\int_{-i\infty}^{+i\infty} {dw''\over 2\,\pi\,i}\,\int_{-\infty}^{+\infty} {d^3 k\over (2\,\pi)^3}\,{e^{i\,k\,R+N\,w''}\over {k^2\over 2}+\left(\sqrt{w''}+{\pi\,\gamma\over 2\,\sqrt{2}}\right)^2}\cr\\
&\cong& {e^{-N\,\Delta e_0(T)/k_B\,T}\over (2\,\pi\,N)^{3/2}}\,\left(1 + {\pi\,\gamma\over 4\,\sqrt{2}}\right)\,\left(1+ {\pi\,\gamma\,N\over 2\,R}\right)\,\,e^{- {R^2\over 2\,N} - {\pi\,\gamma\,R\over 2}}.
\end{eqnarray}
The final result is accurate up to corrections of relative order $1/N$.  All of the shifts of the Laplace variable (renormalization corrections) are combined here into a shift of the bare excitation energy $e_0$:
\begin{equation}
\label{deltae0}
{\Delta e_0(T)\over k_B\,T} \equiv  -\gamma(T)\,{k_{max}\over \pi^2}+ {\pi^2\,\gamma(T)^2\over 8}.
\end{equation}
This expression is apparently the beginning of a series in powers of $\gamma$ that may -- or may not -- be convergent. The main effect of $\Delta e_0(T)$ is to shift $T_0$; and, because we have no {\it a priori} estimate of the unshifted $T_0$, it is easiest to omit this term altogether. On the other hand, if $\gamma(T)$ is large, its temperature dependence will be important in fitting experimental data near $T_0$.  In what follows, I omit $\Delta e_0(T)$, and note simply that Eq.(\ref{Gfinal}) exhibits the expected behavior specified in Section II, specifically, the factor $\exp\,(- \pi\,\gamma\, R/2)$.  

The various shortcomings of this perturbation-theoretic result point to needs for further investigation. The approximation used here is unlikely to be accurate enough for exploring the excitation-chain model quantitatively if the values of $\gamma$ are as large as anticipated.  Also, the calculation properly should include the exclusion effect along with the disorder, instead of dealing with each of these effects separately as if they were decoupled from one another.  And an accurate theory should account more carefully for the molecular-scale features that must play an important role near $T_A$, where the chains become short. What theoretical techniques might be effective for solving the large-$\gamma$ versions of this model? Perhaps diagrammatic techniques can be helpful in computing systematic corrections to the localization factor $\exp\,(-\pi\,\gamma\,R/2)$.  But perturbation expansions are notoriously incapable of predicting some of the most interesting behaviors in systems of this kind. The only possibly relevant, nonperturbative, strong-coupling approach with which I am familiar is my 1966 calculation, in collaboration with Zittartz, of the density of states in an electronic impurity-band tail, where the result has an essential singularity as a function of the disorder strength. \cite{ZL66} That calculation, however, dealt specifically with the statistics of deeply bound states in a white-noise potential, and not the diffusive motion of a particle in that potential.  I have not yet found a way to apply that nonperturbative  method to the present situation.  (See \cite{CATES-BALL88,GOLDSCHMIDT04} for some ideas about non-perturbative calculations in the related context of polymer chains in disordered media.) 

\section{Thermodynamics}

The excitation-chain theory makes no assumptions about the underlying thermodynamic properties of the glassy materials in which the excitations occur.  In this sense, it is closer to the kinetically constrained models of Fredrickson, Andersen, and others \cite{FREDRICKSON85,FREDRICKSON86,KOB93,TONINELLI04} than it is, for example, to Derrida's random-energy model \cite{DERRIDA81} or Wolynes' hypothesis of a random first-order phase transition.\cite{WOLYNES00}  The kinetically constrained models have dynamic properties that look very much like those of glasses -- {\it i.e.} dramatic slowing of relaxation rates similar to that found in the present theory.  One might even postulate that STZ-like defects, containing excess free volume, could serve as the facilitating sites introduced in \cite{FREDRICKSON85}.  (See also the work of Garrahan, Chandler, and coworkers \cite{GARRAHAN-CHANDLER02,GARRAHAN-CHANDLER03,GARRAHAN-CHANDLER05} for a related point of view about the roles of facilitating sites.) Like the present model, the kinetically constrained models require no thermodynamically singular behavior.  It might be interesting to learn whether they would exhibit anything like a Kauzmann phenomenon if they were used in numerical simulations of differential scanning calorimetry experiments.

On the experimental side, decades of careful thermodynamic measurements indicate that the configurational entropy of glassy materials drops linearly toward zero at the Kauzmann temperature $T_K$ -- a behavior that is known to be consistent with equilibrium statistical mechanics only for model systems with long-range and usually built-in random interactions between their constituent elements.  There are striking relations between the Kauzmann thermodynamic phenomenon and the dynamics of real glasses.  The Kauzmann temperature $T_K$ seems to be very close to, and possibly exactly the same as, the Vogel-Fulcher temperature $T_0$.  The Adam-Gibbs theory\cite{ADAM-GIBBS65}, which says that the logarithm of the viscosity is proportional to the inverse of the configurational entropy near $T_0 \cong T_K$, seems to be consistent with a significant body of experimental data.  And there are indications of at least a correlation, and perhaps a direct proportionality between the measured jump in the configurational specific heat at the glass transition and the dynamic fragility in a wide range of glassy materials.\cite{WOLYNES00} I find myself deeply puzzled by this situation.  It is hard for me to believe that long-range, mean-field models can accurately predict the dynamical behavior of glasses consisting of small molecules with short-range interactions.  Moreover, as mentioned in the Introduction, I am not convinced that the connection between dynamics and thermodynamics has properly been established for  mean-field models that exhibit a thermodynamic Kauzmann phenomenon.  On the other hand, the thermodynamic observations must be taken very seriously.

The excitation-chain theory provides one clue that may point toward a resolution of the dilemma.  This purely dynamic mechanism predicts a length scale, $R^*(T)$ in Eq.(\ref{R*}),  that diverges like $(T-T_0)^{-1}$ near $T_0$ and vanishes at $T_A$.  An isolated region of the system that is smaller than $R^*$ would not be large enough to sustain critically long excitation chains; therefore its constituents would be frozen and unable to make thermally activated transitions between different inherent states.  (The molecules in such a region, of course, would continue to undergo rapid thermal fluctuations within their cages; but these fluctuations would be unable to activate configurational rearrangements.)  Regions larger than $R^*$, however, would support critical excitations and be able to change their sizes and shapes via molecular rearrangements, albeit very slowly at low temperatures.  If such regions exist in any meaningful way, either as observably bounded domains or simply as slowly fluctuating volumes in which the correlations are extremely long-lasting, then their characteristic sizes would be proportional to $R^*$. Smaller regions would be frozen and would be able to grow only by coming into contact with larger, unfrozen regions; the latter regions would shrink to increase the entropy of the system as a whole. 

The question of whether spatial heterogeneities occur in glass forming materials seems not yet to be clearly resolved. See reviews by Sillescu \cite{SILLESCU99} and Ediger \cite{EDIGER00}, who argue in favor of heterogeneity. In a recent paper, Berthier {\it et al} \cite{BERTHIER05} point out that heterogeneity in glasses can be detected by measuring multi-point dynamic susceptibilities, and show experimental data indicating that heterogeneities exist.  Shi and Falk \cite{SHI05}, in STZ-related molecular dynamics simulations, have seen spatial patterns apparently similar to what I propose here. They find that their two dimensional Lennard-Jones glass, when annealed, consists primarily of domains in which the molecules are strongly correlated in low-energy configurations, and that these strongly correlated domains are separated by interfaces containing higher energy defects.  

To solve the thermodynamic puzzle, I postulate that the only unfrozen configurational degrees of freedom of this system exist on the boundaries of strongly correlated regions of size $R^*$.  In other words, I conjecture that the configurational entropy per molecule of the system as a whole is proportional to the surface-to-volume ratio, $1/R^*$, of such regions.  Specifically,
\begin{equation}
\label{entropy}
s_c(T)\approx {s_0\over R^*(T)} = {1\over 2}\,\left({6\over \pi}\right)^{1/2}\,{s_0\,\nu\,(T-T_0)\over (\gamma_0\,T_0\,T_{int})^{1/2}},
\end{equation}
where $s_c(T)$ is the average configurational entropy per molecule in units of $k_B$, and $s_0$ is the configurational entropy per unfrozen molecule.  The entropy $s_0$ is a measure of how free the boundary molecules are to rearrange their positions, orientations and, in the case of complex molecules, their internal configurations -- freedoms that presumably are lacking inside the frozen regions. 

Here, and in what follows, I assume that $T$ is sufficiently close to $T_0$ that the long-chain approximation is accurate, and that the super-Arrhenius part of Eq.(\ref{DeltaGnew}) dominates the Arrhenius part, $\Delta G_{\infty}$. Equation (\ref{entropy}) should not be taken literally out to temperatures in the neighborhood of $T_A$, where -- as in the case of $\Delta G^*(T)$ -- some correction for vanishingly short chains will be needed, and where $s_c(T)$ must join smoothly to the configurational entropy of the liquidlike state. 

Equation (\ref{entropy}) implies that the configurational entropy vanishes linearly in $T$ at the Kauzmann temperature $T_K$, and that $T_K = T_0$.  The combination of Eqs.(\ref{DeltaGnew}) and (\ref{entropy}) yields
\begin{equation}
{\Delta G^*(T)\over k_B\,T} \approx {\pi\,\gamma_0\,s_0\over 3\,s_c(T)},
\end{equation}
which is essentially the Adam-Gibbs formula near $T_0$.  

Definitions of the glass temperature $T_g$ generally have the form
\begin{equation}
{\Delta G^*(T_g)\over k_B\,T_g} \approx 4\,\left({\pi\over 6}\right)^{3/2}\,{\gamma(T_g)^{3/2}\,(T_g\,T_{int})^{1/2}\over \nu\,(T_g-T_0)} = \lambda_g,
\end{equation}
where $\lambda_g$ is a large number of order $30$ or so, chosen roughly to represent the observable limits of long relaxation times or high viscosities.  Then $T_g\cong T_0$, and the fragility $m$ \cite{ANGELL95} is 
\begin{equation}
m \equiv - T\,{\partial\over \partial T}\,\left({\Delta G^*(T)\over k_B\,T}\right)_{T=T_g} \approx {1\over 4}\,\left({6\over \pi}\right)^{3/2}\,{\nu\,\lambda_g^2\over \gamma_0}\,\left({T_g\over \gamma_0\,T_{int}}\right)^{1/2}.
\end{equation} 
Thus the excitation-chain theory implies that glasses are fragile when $T_{int} \ll T_0$, and/or when the disorder strength $\gamma_0$ is small.  Returning to the thermodynamic formula, Eq.(\ref{entropy}), we find that the jump in the specific heat at $T_g \cong T_0$ is
\begin{equation}
\Delta c_p = \left(T\,{\partial s_c\over \partial T}\right)_{T=T_0} \approx {2\,s_0\,\gamma_0\,m\over \lambda_g^2}.
\end{equation}
Here we recover the conjectured proportionality between $\Delta c_p$ and $m$, but with two material-specific parameters, $s_0$ and $\gamma_0$, that might account for the observed scatter in the experimental data.  According to its definition, $s_0$ should scale with the ``bead'' number of the molecules, which ordinarily is factored out in obtaining the linear relation between $\Delta c_p$ and $m$.  (See \cite{WOLYNES00}.)  Because it is basically a geometrical quantity, not involving energy scales, $\gamma_0$ may be roughly a constant, of order unity, for most glassy materials.  This analysis also implies that
\begin{equation}
R^*(T_g)\approx \left({3\over \pi}\right)\,{\lambda_g\over \gamma_0}.
\end{equation}
Thus the critical length scale $R^*$ at the glass temperature is predicted to be about $30$ molecular spacings, independent of the fragility.  This prediction seems to be qualitatively consistent with results shown by Berthier {\it et al.} \cite{BERTHIER05}; but those authors report a substantially smaller length scale.  

Clearly, this thermodynamic analysis is incomplete and highly speculative.  The excitation-chain theory and its proposed extension to thermodynamics both need to be explored further and tested experimentally.  We also need to develop a theory of the crossover between super-Arrhenius and Arrhenius behavior at $T_A$, and to understand how the present solidlike formulation crosses over to liquidlike mode-coupling theories at higher temperatures.\cite{GOTZE91,GOTZE92} And we have yet to address the important issue of stretched-exponential relaxation.  (See \cite{LEMAITRE02} for an idea about how stretched exponentials might appear in STZ theory.)  

However, the deepest theoretical uncertainty is the meaning of Eq.(\ref{entropy}).  If correct, this relation implies that ergodicity is broken not just below the Kauzmann temperature but, to a continuously increasing extent as $T$ falls below $T_A$, throughout the super-Arrhenius region $T_0 < T < T_A$.  The language that I have used to support this conjecture is at best suggestive, and is far from being a systematic derivation of that equation.  A list of unanswered questions makes the uncertainties abundantly clear.  Is the domain structure to be taken literally, or is it just a way of talking about long-lasting correlations?  Might there be some kind of long-range order -- orientational or perhaps something even more subtle -- inside the domains?  Might the excitation chains occur predominantly in the more highly disordered boundary regions and, if so, might we need to return to the two-dimensional picture proposed in \cite{LL05}?  Is a ``frozen'' region permanently frozen?  Presumably not, because the domain boundaries must diffuse at something like the Vogel-Fulcher relaxation rate.  On the other hand, if all of the dynamically accessible states of the system consist of frozen domains separated by unfrozen boundaries, then Eq.(\ref{entropy}) might be justified as the appropriate statistical average. How might such a picture be made into a quantitative, predictive theory?  What theoretical tools might be useful?

\begin{acknowledgments}
This research was supported by U.S. Department of Energy Grant No. DE-FG03-99ER45762. I would like especially to thank Anael Lemaitre for his collaboration on the preceding paper and for  help in analyzing experimental data.  I also would like to thank  Jean-Louis Barrat, Jean-Philippe Bouchaud, Glenn Fredrickson, and Swagatam Mukhopadhyay for discussions and assistance.  
\end{acknowledgments}


\begin{thebibliography}{99}

\bibitem{LL05} J.S. Langer and A. Lemaitre, Phys. Rev. Lett. {\bf 94}, 175701 (2005).

\bibitem{JAMMING} {\it Jamming and Rheology}, A.J. Liu and S.R. Nagel, eds.,(Taylor and Francis, London and New York, 2001).

\bibitem{FL98} M.L. Falk and J.S. Langer, Phys. Rev. E {\bf 57}, 7192 (1998).

\bibitem{FLP04} M.L. Falk, J.S. Langer, and L. Pechenik, Phys. Rev. E. {\bf 70}, 011507 (2004).

\bibitem{JSL04} J.S. Langer, Phys. Rev. E {\bf 70}, 041502 (2004).

\bibitem{JSL69} J.S. Langer, Ann. Phys. {\bf 54}, 258 (1969).

\bibitem{KIVELSON95} D. Kivelson, S.A. Kivelson, X. Zhao, Z. Nussinov, and G. Tarjus, Physica {\bf A 219}, 27 (1995).

\bibitem{WOLYNES89} T.R. Kirkpatrick, D. Thirumalai, and P.G. Wolynes, Phys. Rev. A {\bf 40}, 1045 (1989).

\bibitem{WOLYNES00} X. Xia and P.G. Wolynes, Proc. Natl. Acad. Sci. U.S.A. {\bf 97}, 2990 (2000); Phys. Rev. Lett. {\bf 86}, 5526 (2001).

\bibitem{WOLYNES05} S.M. Bhattacharyya, B. Bagchi, and P.G. Wolynes, arXiv:cond-mat/0505030 (2005).

\bibitem{FLORY53} P.J. Flory, {\it Principles of Polymer Chemistry}, Cornell University Press, Ithaca, NY (1953).

\bibitem{ADAM-GIBBS65} G. Adam and J.H. Gibbs, J. Chem. Phys. {\bf 43}, 139 (1965).

\bibitem{TARJUS-KIVELSON} G. Tarjus and D. Kivelson, "The viscous slowing down of supercooled liquids and the glass transition: phenomenology, concepts, and models," in \cite{JAMMING}. 

\bibitem{BOUCHAUD04} J.-P. Bouchaud and G. Biroli, J. Chem. Phys. {\bf 121}, 7347 (2004).

\bibitem{GOLDSTEIN69} M. Goldstein, J. Chem. Phys. {\bf 51}, 3728 (1969).

\bibitem{STILLINGER-WEBER82} F.H. Stillinger and T.A. Weber, Phys. Rev. A {\bf 25}, 978 (1982).

\bibitem{STILLINGER88} F.H. Stillinger, J. Chem. Phys. {\bf 88}, 7818 (1988).

\bibitem{COHEN-GREST} M. H. Cohen and G. Grest, Phys. Rev. B {\bf 20} (3), 1077 (1979).

\bibitem{SPAEPEN77} F.Spaepen, Acta Metall. {\bf 25}(4), 407 (1977).

\bibitem{ARGON79} A.S. Argon, Acta Metall. {\bf 27}, 47 (1979).

\bibitem{DUINE} P.A. Duine, J. Sietsma, and A. Van den Beukel, Acta Metall. Mater. {\bf 40}, 743 (1992). 

\bibitem{TUINSTRA} P. Tuinstra, P. Duine, J. Sietsma, and A. Van den Beukel, Acta Metall. Mater. {\bf 43}, 2815 (1995). 

\bibitem{DEHEY} P. De Hey, J. Sietsma, and A. Van den Beukel, Acta Mater. {\bf 46}, 5873 (1998).

\bibitem{WLJ05} M.D. Demetriou and W.L. Johnson, Scripta Materialia {\bf 52}, 833 (2005).

\bibitem{GLOTZER99} C. Donati, S. Glotzer, P. Poole, W. Kob, and S. Plimpton, Phys. Rev. E {\bf 60}, 3107 (1999).

\bibitem{GLOTZER00a} S.C. Glotzer, J. Non-Crystalline Solids {\bf 274}, 342 (2000).

\bibitem{GLOTZER00b} T.B. Schr\o der, S. Sastry, J.P. Dyre, and S.C. Glotzer, J. Chem. Phys. {\bf 112}, 9834 (2000).

\bibitem{EDWARDS65} S.F. Edwards, Proc. Phys. Soc. {\bf 85}, 613 (1965).

\bibitem{KTZ+96} D. Kivelson, G. Tarjus, X. Zhao, and S.A. Kivelson, Phys. Rev. E {\bf 53(1)} 751 (1996).

\bibitem{GELFAND-YAGLOM60} I.M. Gel'fand and A.M. Yaglom, J. Math. Phys. {\bf 1}, 48 (1960).

\bibitem{CATES-BALL88} M.E. Cates and R.C. Ball, J.Physics (France) {\bf 89}, 2435 (1988).

\bibitem{GOLDSCHMIDT04} Y.Y. Goldschmidt and Y. Shiferaw, arXiv:cond-mat/0411195, to be published in "Statistics of Linear Polymers in Disordered Media", edited by B.K. Chakrabarti, (Elsevier 2005). 

\bibitem{KOHN-LUTTINGER57} W. Kohn and J.M. Luttinger, Phys. Rev. {\bf 108}, 590 (1957).

\bibitem{MAHAN} G.D. Mahan, {\it Many-Particle Physics} (Plenum Press, New York, 1981).


\bibitem{ZL66} J. Zittartz and J.S. Langer, Phys. Rev. {\bf 148}, 741 (1966).

\bibitem{FREDRICKSON85} G.H. Fredrickson and H.C. Andersen, J. Chem. Phys. {\bf 83}, 5822 (1985).

\bibitem{FREDRICKSON86} G.H. Fredrickson and S.A. Brawer, J. Chem. Phys. {\bf 84}, 3351 (1986).

\bibitem{KOB93} W. Kob and H.C. Andersen, Phys. Rev. E {\bf 48}, 4364 (1993).

\bibitem{TONINELLI04} C. Toninelli, G. Biroli, and D.S. Fisher, Phys. Rev. Lett. {\bf 92}, 188504 (2004).

\bibitem{DERRIDA81} B. Derrida, Phys. Rev. B {\bf 24}, 2613 (1981).

\bibitem{GARRAHAN-CHANDLER02} J.P. Garrahan and D. Chandler, Phys. Rev. Lett. {\bf 89}, 03574 (2002).

\bibitem{GARRAHAN-CHANDLER03} J.P. Garrahan and D. Chandler, Proc. Natl. Acad. Sci. {\bf 100}, 9710 (2003).

\bibitem{GARRAHAN-CHANDLER05} M. Merolle, J.P. Garrahan and D. Chandler, Proc. Natl. Acad. Sci. {\bf 102}, 10837 (2005). 

\bibitem{ANGELL95} C.A. Angell, Science {\bf 267}, 1924 (1995).

\bibitem{SILLESCU99} H. Sillescu, J. Non-Crystalline Solids {\bf 243}, 81 (1999).

\bibitem{EDIGER00} M.D. Ediger, Ann. Rev. Phys. Chem. 2000.51:99-128 (2000).

\bibitem{BERTHIER05} L. Berthier, G. Biroli, J.-P. Bouchaud, L. Cipelletti, D. El Masri, D.L'Hote, F. Ladieu, and M. Pierno, Science {\bf 310}, 1797 (2005).   

\bibitem{SHI05} Y. Shi and M.L. Falk, Phys. Rev. Lett. {\bf 95}, 095502 (2005), and unpublished results. 

\bibitem{GOTZE91} W. G\"otze, in {\it Liquids, Freezing and Glass Transition}, ed. J.-P. Hansen, D. Levesque and J. Zinn-Justin (North Holland, Amsterdam, 1991), p. 289. 

\bibitem{GOTZE92} W. G\"otze and L. Sj\"ogren, Rep. Prog. Phys. {\bf 55}, 241 (1992).


\bibitem{LEMAITRE02} A. Lemaitre, arXiv:cond-mat/0206417.

\end{thebibliography}
\end{document}